\documentclass[12pt,preprint]{aastex}

\usepackage{graphicx}
\usepackage{epstopdf}

\shorttitle{The Case Against Dark Matter and Modified Gravity}
\shortauthors{Christodoulou \& Kazanas}

\begin{document}


\title{The Case Against Dark Matter and Modified Gravity: \\
Flat Rotation Curves Are a Rigorous Requirement \\
in Rotating Self-Gravitating Newtonian Gaseous Disks}


\author{Dimitris M. Christodoulou}
\affil{Department of Mathematical Sciences, \\
and Lowell Center for Space Science and Technology, \\
University of Massachusetts Lowell, \\
Lowell, MA 01854 \\
\email{dimitris\_christodoulou@uml.edu}}

\and

\author{Demosthenes Kazanas}
\affil{NASA Goddard Space Flight Center, \\
Laboratory for High-Energy Astrophysics, \\
Code 663, Greenbelt, MD 20771 \\
\email{demos.kazanas@nasa.gov}}




\begin{abstract}
By solving analytically the various types of Lane-Emden equations with rotation,
we have discovered two new coupled fundamental properties of rotating, self-gravitating, gaseous disks in equilibrium: Isothermal disks must, on average, exhibit strict power-law density profiles in radius $x$ on their equatorial planes
of the form $A x^{k-1}$, where $A$ and $k-1$ are the integration constants;
 and ``flat'' rotation curves precisely such as those observed in spiral galaxy disks. 
Polytropic disks must, on average, exhibit strict 
density profiles of the form $\left[\ln(A x^k)\right]^n$, where 
$n$ is the polytropic index; and ``flat'' rotation curves described
by square roots of 
upper incomplete gamma functions.  By ``on average,'' we mean that,
irrespective of the chosen boundary conditions, the actual profiles must oscillate around and remain close to the strict mean profiles of the analytic singular equilibrium solutions. 
We call such singular solutions the ``intrinsic'' solutions of the differential equations because
they are demanded by the second-order equations themselves with no regard to the Cauchy problem. The results are directly applicable to gaseous galaxy disks
that have long been known to be isothermal and to protoplanetary disks during the extended
isothermal and adiabatic phases of their evolution. In galactic gas dynamics, they have the potential to resolve
the dark matter--modified gravity controversy in a sweeping manner, as they render both of these hypotheses unnecessary. In protoplanetary disk research, they provide observers with a powerful new probing tool, as they predict a clear and simple connection between the radial density profiles and the rotation curves of self-gravitating disks in their very early (pre-Class 0 and perhaps the youngest Class 0 Young Stellar Objects) phases of evolution. 
\end{abstract}





\section{Introduction}\label{intro}

A large number of observations, mostly in the 21 cm emission line of neutral hydrogen, 
have firmly established that the rotation curves of spiral galaxy disks do not exhibit
a Keplerian falloff, in fact, most of them remain flat or slightly increasing as far away from the centers as they can be observed \citep{fre70,rog72,rob73,bos78,rub80,rft80,bos81a,bos81b,rft82,bah85,van86,ken87,beg87,per88,beg89,per90,car90,cas91,bro92,per95,per96,sal97}.
The radial scales over which the neutral hydrogen disks can be observed reach out to $\sim$100~kpc in the largest spiral galaxies and since the rotation curves remain flat, it was postulated by many researchers that some unseen extended mass distribution ought to exist all the way out to hundreds of kiloparsecs from the galaxy centers. Thus the Dark Matter Hypothesis was born, and soon the news leaked out to the rest of the physics community and intensive and extensive searches for dark matter particles and fields boomed into existence. On the other hand, some researchers who certainly felt uncomfortable with this new ``aetherial'' hypothesis proposed that Newtonian gravity should instead be modified at galactic scales and beyond, in order to solve the problem of the fast rotation of HI galaxy disks \citep{mil83a,mil83b,mil83c,toh83,fel84,san84,san86,man89,chr91,man11,man12}.

The gas in spiral galaxies is distributed in centrally concentrated, vertically thin disks.
For this reason, it was expected that the rotation curves had to turn over at some intermediate radius and begin a decline that would be indicative of the absence of substantial amounts of matter at large radii. This view about the luminous matter is nowadays considered so settled
and clear that it has made its way into introductory Astronomy textbooks that compare and contrast the kinematics of spiral galaxies to the kinematics observed in our solar system
(the Keplerian falloff mentioned above).
In this work, we show that this elementary perception is quite naive and totally wrong
because it ignores the influence (in fact, the dominance) of pressure and enthalpy gradients in 
self-gravitating gaseous disks.
Personally, we believe that it amounts to a blunder because, before the results described below and even back in the 1980s when all this was unfolding, we had important clues that pointed out the importance of gas pressure in determining the equilibrium structures of gaseous disks.
For instance, the sound-crossing time at 10~kpc in a 10$^4$~K cold galaxy disk is 1~Gyr, a value that lies to within 1.5-3 of the rotation timescales at 10~kpc in all galaxies with rotation speeds of 100-200~km~s$^{-1}$. 
\cite{gal84} noted that the star formation histories of a sample of irregular and spiral galaxies did not indicate the presence of dark matter 
in the low-mass galaxies of the sample.
But the most obvious clue was the existence of the \cite{mes63} disk,
a centrally concentrated potential-density pair with a flat rotation curve
(see also Jalali \& Abolghasemi 2002). The Mestel disk
has always been considered just a toy model \citep{bin87}
and the situation did not improve when \cite{sch12}
showed that the finite Mestel disk requires {\it significantly less mass} to produce the flat rotation curve. The main argument against the universal adoption of the Mestel disk has been the absence of a physical law or reason that would make galaxy disks assume this specific surface density profile and rotation
law.

The above argument is not based on solid reasoning. 
When nature shows us that she has widely adopted 
a specific property (the flat rotation curves in galaxy disks), Aristotelian Logic dictates that we should search for a new law or reason, in order to understand the universality of this property and establish its physical meaning;
not to create ghosts (particles and fields), aethers, and new forces
that effectively facilitate our aversion to confronting the facts.

We do not claim that the \cite{mes63} disk is the answer to establishing the universality of flat rotation curves in galaxy disks; only that it has always been a telling clue that gravity does not pull the strings and is not in control in gaseous self-gravitating disks. Furthermore, 
we have solved the full Newtonian problem and we now know precisely 
how such universal rotation curves emerge in spiral galaxy disks. 
The resolution of this ubiquitous problem is the subject of this paper. 
Before we can delve
into the physics of the problem, we need to correct some common misconceptions that appear in the theory of second-order differential equations and which also have made their way into the textbooks. We do so in \S~\ref{math}. Then, in \S~\ref{iso}, we revisit the theory of rotating Newtonian isothermal gaseous-disk equilibrium models and we calculate analytically the mean shapes of their density profiles and their rotation curves. The results match precisely the shapes of the rotation curves of spiral galaxy disks with no additional assumptions of any kind. So these results make a strong case against both dark matter and modified gravity and their implications have far-reaching consequences all the way to cosmology.
For completeness, we describe in \S~\ref{poly} polytropic models that also demand monotonically increasing rotation curves because they are subject to the same physical principles. These models are also applicable to very young protoplanetary disks (certainly to pre-Class 0 disks and possibly to the youngest Class 0 non-Keplerian disks).
Finally, we conclude with a discussion of all the pertinent issues and our results in \S~\ref{discussion}.

\section{Second-Order Differential Equations and the Cauchy Problem}\label{math}

In mathematical physics, the trivial solutions of the various second-order differential equations 
are commonly ignored as being uninteresting; and too much attention is paid to 
the Cauchy problem in determining arbitrary constants as opposed to the internal properties of the equations themselves that have no regard for externally imposed conditions of any type. 
Both of these practices are damaging as they work to hinder our efforts toward solving the physical problems described by the differential equations in the first place.
Such practices are relevant to all linear and nonlinear 
second-order equations of physics, so we can discuss and clarify
the various issues involved by using any well-known equation. 
We choose to make use of the
Bessel differential equation in this section.   

\begin{figure}
\epsscale{.80}
\plotone{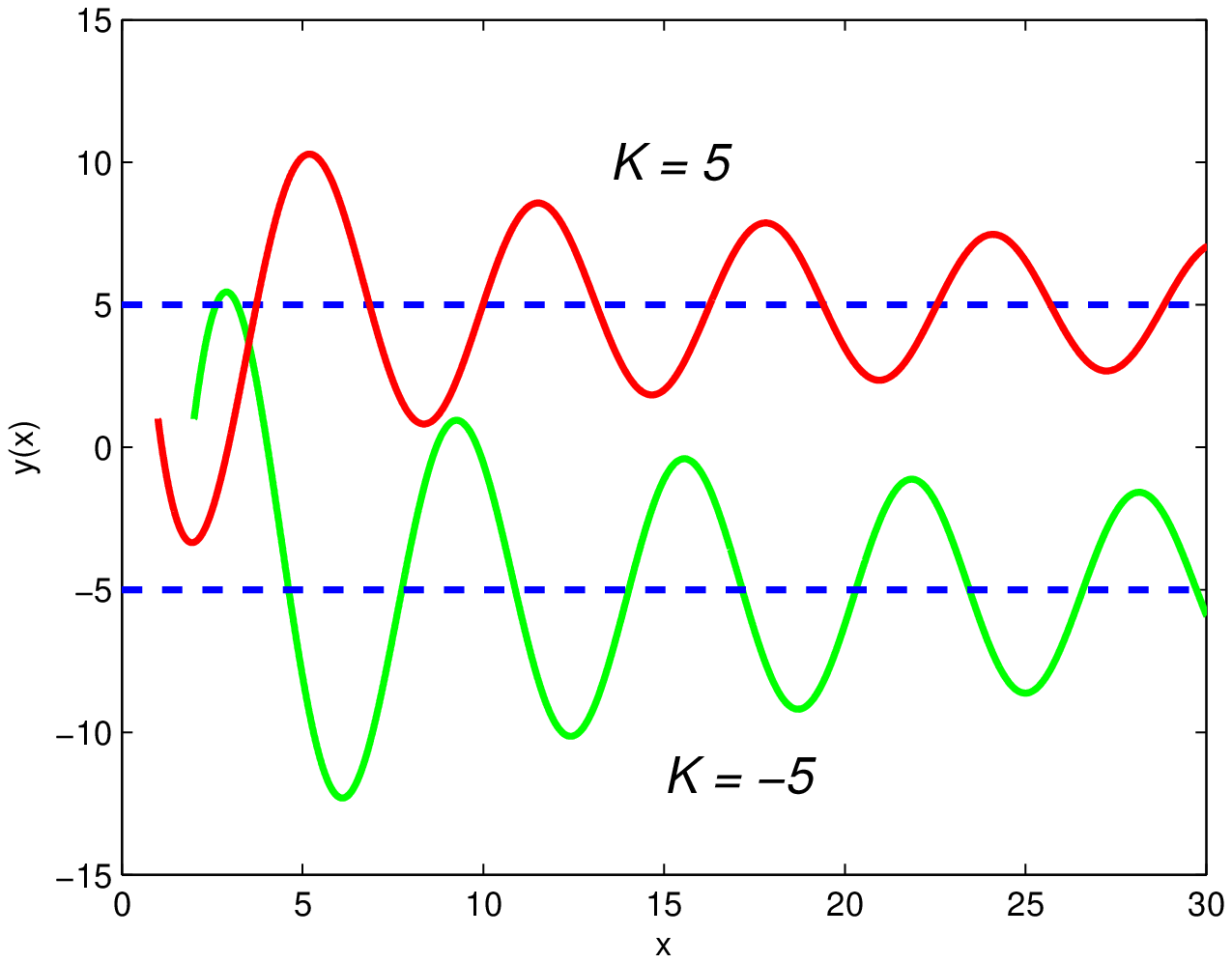}
\caption{Numerical solutions of the $m=0$ inhomogeneous 
Bessel differential equation~(\ref{bessel2}) subject to the
following boundary conditions:
in the $K=5$ case, we use $y(1)=1$ and $y'(1)=-10$;
in the $K=-5$ case, we use $y(2)=1$ and $y'(2)=10$. 
In both cases, the regular solutions are forced to oscillate and stay near
the dominant intrinsic solutions $y=K$ (dashed lines).
\label{fig1}}
\end{figure}

The Bessel equation \citep{wat22}
\begin{equation}
y'' + \frac{1}{x}y' + \left(1 - \frac{m^2}{x^2}\right)y = 0, \ \ (m = {\rm const.}),
\label{bessel1}
\end{equation}
has regular solutions that are called Bessel functions of order $m$ and a trivial solution $y=0$.
The trivial solution is not singular as it can be obtained from the regular solutions by an
appropriate choice of the arbitrary constants. Nevertheless, its name indicates that $y=0$
is of no interest at all. It is also well-known that the Bessel functions all oscillate about the $x$-axis
\citep{wat22}, but this statement is grossly inaccurate and obscures the truth: the regular solutions oscillate about the trivial solution which just happens to coincide with the $x$-axis in this case.
We demonstrate this important point by solving numerically an inhomogeneous $m=0$ Bessel equation of the form
\begin{equation}
y'' + \frac{1}{x}y' + y = K,  \ \ (K = {\rm const.}),
\label{bessel2}
\end{equation}
along with nonsingular boundary conditions that attempt to initially push the regular solutions away from the new trivial solutions $y=K$: 
$y(1)=1$, $y'(1)=-10$ for $K=5$;
and $y(2)=1$, $y'(2)=10$ for $K=-5$. 
The results are shown in Fig.~\ref{fig1}.
Both regular solutions have nothing to do with the $x$-axis; instead, they turn around and settle into oscillations that clearly occur
about the new trivial solutions $y=K$.  
This behavior can be demonstrated for all
linear second-order equations of mathematical physics
with oscillatory regular solutions \citep{chr15} 
and for (non)linear equations of the Lane-Emden type \citep{rob68,sch86,sch95,chr07}.
The lesson to be learned is that the so-called trivial solutions of second-order equations
are not at all trivial. They are in fact favored by the differential equations 
themselves which have no regard for the externally imposed boundary conditions.
Thus, we will heretofore call these solutions the {\it intrinsic solutions} that are
preferred and demanded by the equations themselves, irrespective of the externally imposed Cauchy problem.

When the Cauchy problem is solved, as in Fig.~\ref{fig1}, the externally
imposed boundary conditions are usually at odds with the underlying equation
and the regular solutions cannot match the favored intrinsic solution.
As a result, the regular solutions are forced by the equation itself to oscillate about the intrinsic
solution as soon as they intersect this favored solution the first time. Thus, the intrinsic solutions
act as attractors of the regular solutions which, in turn, are forced to always stay near and around
the more dominant intrinsic solutions.  We view this behavior
as a triumph of the differential equation (and its intrinsically favored solution) over the Cauchy problem
(and the particular solution it strives to produce). 

This striking behavior remains intact in at least some nonlinear second-order equations.
Very clear examples in which rotation is involved can be found in \cite{sch95} and \cite{chr07}. 
Here we provide two additional nonlinear examples of the dominance of intrinsic solutions 
over regular solutions drawn from Lane-Emden equations \citep{lan69,emd07}
in the absence of rotation.
The isothermal Lane-Emden equations take the form
\begin{equation}
y'' + \frac{D-1}{x}y' + e^y = 0, 
\label{leiso0}
\end{equation}
where $D$ is the dimensionality of space ($D=3$ in spherical coordinates and
$D=2$ in cylindrical coordinates). Singular and regular solutions have been obtained
in many applications \citep{sto63,ost64,sch86,bin87,chr07} and they are all 
nonoscillatory.\footnote{The periodic solution found 
by \cite{sch67} describes a cartesian slab and not a disk or cylinder.}
The reason for this is quite obvious: eq.~(\ref{leiso0}) does not have an 
intrinsic solution because $e^y\neq 0$.
Why the latter condition precludes an intrinsic solution
will become clear in the next section, where we describe a procedure for obtaining intrinsic solutions.

\begin{figure}
\epsscale{.80}
\plotone{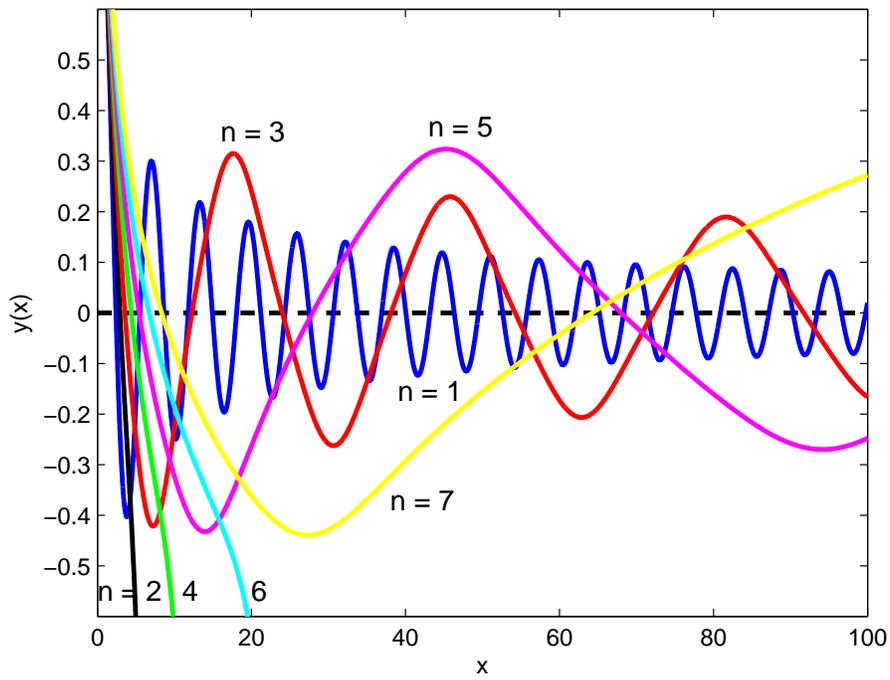}
\caption{Numerical solutions of the cylindrical ($D=2$) polytropic 
Lane-Emden equation~(\ref{lepoly0}) for integer values of $n$ from
1 to 7 and subject to the usual boundary conditions
$y(0)=1$ and $y'(0)=0$. Only the solutions for odd values of $n$ oscillate
about the intrinsic solution $y=0$ (dashed line).
\label{fig1b}}
\end{figure}

\begin{figure}
\epsscale{.80}
\plotone{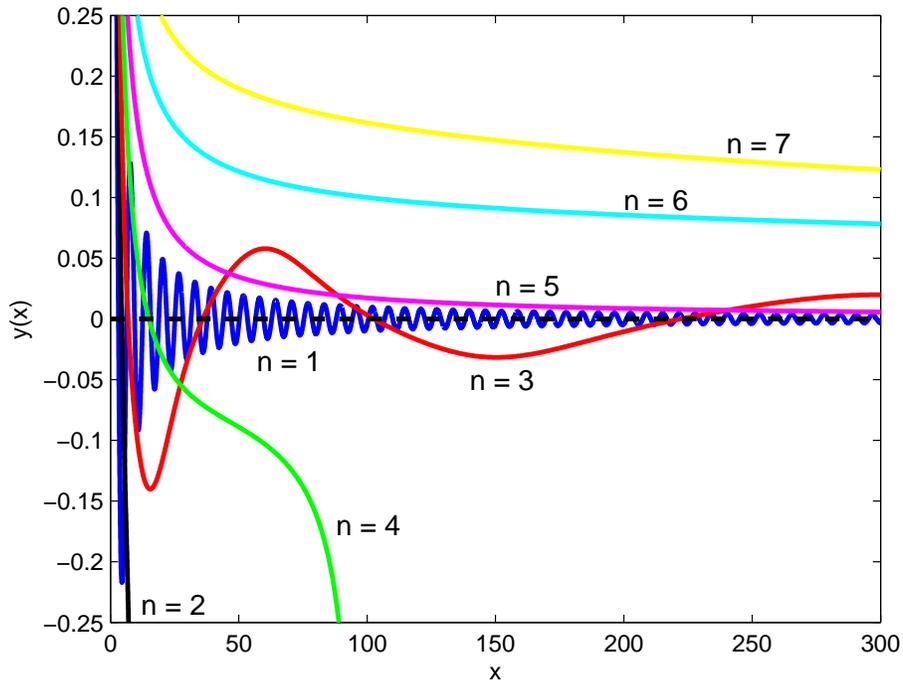}
\caption{Numerical solutions of the spherical ($D=3$) polytropic 
Lane-Emden equation~(\ref{lepoly0}) for integer values of $n$ from
1 to 7 and subject to the usual boundary conditions
$y(0)=1$ and $y'(0)=0$. Only the solutions for $n=1$ and $n=3$ oscillate
about the intrinsic solution $y=0$ (dashed line).
\label{fig1c}}
\end{figure}

In stark contrast, the polytropic Lane-Emden equations take the form
\begin{equation}
y'' + \frac{D-1}{x}y' + y^n = 0, 
\label{lepoly0}
\end{equation}
where $n>0$ is the polytropic index, and it possesses the intrinsic 
solution $y=0$. Although few analytic solutions are known \citep{cha39}, numerical integrations show that, depending on $n$, this equation has both oscillatory and nonoscillatory solutions. 
For eq.~(\ref{lepoly0}), we have derived a precise criterion for the existence 
of oscillatory solutions and we will describe it in a future publication. 
This criterion predicts that for $D=2$ (cylindrical form), all solutions with odd integer 
$n$-values are oscillatory; while for $D=3$ (spherical form), only the $n=1$ and $n=3$ integer-$n$ solutions are oscillatory. Numerical integrations 
(using the physical boundary conditions $y(0)=1$, $y'(0)=0$)
easily confirm these results (see Figs.~\ref{fig1b} and~\ref{fig1c}). The reason for the existence of nonoscillatory
solutions is that for the corresponding choices of $n$, the differential
equation is not a harmonic oscillator \citep{chr15}. This is also true for the modified
Bessel equation \citep{wat22}
that is known to possess only nonoscillatory solutions.
Its real solutions cannot be oscillatory\footnote{We note
in passing that, as a result of the substitution $x\to ix$ that produces the
modified Bessel equation \citep{wat22}, its solutions 
are oscillatory about the imaginary axis
in the comlex plane.}  
and they are then prohibited from intersecting the intrinsic solution 
more than once \citep{chr15}.

\section{Isothermal Self-Gravitating Newtonian Gaseous Disks}\label{iso}

In what follows, we use the arbitrary scaling constants $R_o$ and $\rho_o$
to normalize the disk radius $R$ and density $\rho(R)$, respectively.
We thus define the dimensionless radius $x\equiv R/R_o$ and density
$\tau(x)\equiv \rho(R)/\rho_o$.
Velocities $V(R)$ are also normalized consistently by the constant
$V_o = R_o\sqrt{4\pi G\rho_o}$,
where $G$ is the Newtonian gravitational constant,
in which case we also define the dimensionless rotation 
velocity $v(x)\equiv V(R)/V_o$. The same scaling also applies to the sound speed $C_o$ of the gas
which in this section is a constant, i.e., the dimensionless
sound speed is $c_o\equiv C_o/V_o$.

The cylindrical isothermal Lane-Emden equation \citep{lan69,emd07}
with rotation can then be written in dimensionless form as
\begin{equation}
c_o^2\cdot\frac{1}{x}\frac{d}{dx}x\frac{d}{dx}\ln\tau + \tau =
\frac{1}{x}\frac{dv^2}{dx} \ .
\label{leiso1}
\end{equation}
This equation describes the radial ($x$) equilibrium of a rotating, 
self-gravitating, gaseous disk or cylinder in which the gas obeys the 
isothermal equation of state $p(x) = c_o^2\tau(x)$, where $p$ 
is the dimensionless pressure of the gas.
Eq.~(\ref{leiso1}) is valid exactly for
infinite cylinders and to a high degree of approximation in the equatorial (symmetry) planes of disks (see the Appendix).
This latter point has been demonstrated convincingly by the calculations of
\cite{hay82}, \cite{sch86}, and \cite{nar90}. In particular, the last two investigations of finite disks uncovered equatorial density profiles that were
strictly oscillatory under proper boundary conditions, just as was predicted by the analysis of \S~\ref{math} above.
 
\cite{hay82} and \cite{schm86} studied also the stability
of such equilibria and found that, except for the very flattened disks and the nearly spherical configurations, the intermediate models are stable. The very flattened disks of \cite{hay82}, in particular, were unstable to ring formation that causes their equatorial power-law density profiles to become oscillatory,
in agreement with the numerical solutions of eq.~(\ref{leiso1})
obtained by \cite{chr07}.  

Despite the exact analytic results of the researchers quoted above, an objection has been raised over the years concerning the validity of using cylindrical coordinates to study axisymmetric, vertically thin disks rather than just infinite cylinders; and this is also the major sticking point in the present work, so it needs to be addressed in detail. We defer the analysis 
of the Lane-Emden equations for vertically thin disks to an Appendix 
because it is much easier to follow the derivations after the intrinsic analytic solution has been obtained for cylinders as follows.

\cite{chr07} described a procedure for obtaining the intrinsic solution of
eq.~(\ref{leiso1}): If we equate the last two terms:
\begin{equation}
\tau(x) = \frac{1}{x}\frac{dv^2(x)}{dx},
\label{rhoiso1}
\end{equation}
then this is an intrinsic solution provided that the rest of the equation 
(the radial variation of the logarithmic gradient of the enthalpy) 
vanishes:
\begin{equation}
\frac{d}{dx}x\frac{d}{dx}\ln\tau(x) = 0.
\label{rhoiso2}
\end{equation}
Eqs.~(\ref{rhoiso1}) and~(\ref{rhoiso2}) form a system
in which $v(x)$ is totally dependent on $\tau(x)$. 
First we solve eq.~(\ref{rhoiso2}) to
obtain the radial density profile:
\begin{equation}
\tau(x) = A x^{k-1}, \ \ (A, k = {\rm const.}),
\label{rhoiso3}
\end{equation}
and then we solve eq.~(\ref{rhoiso1}) to determine the rotation curve
of the intrinsic solution:
\begin{equation}
v(x) = \sqrt{A g(x) + B}, \ \ (B = {\rm const.}),
\label{rhoiso4}
\end{equation}
where
\begin{equation}
g(x) \ \equiv \  \left\{ \begin{array}{cc} 
         x^{k+1}/(k+1) \ , & \ {\rm if} \ \ \ k \neq -1 \\
         \ln x \ , \ \ \ \ \ \ \ \ \ \ \ & \ {\rm if} \ \ \ k = -1 
         \end{array} \right. \ .
\label{g}
\end{equation}
The solution contains 3 free parameters, the integration constants
$A$, $B$, and $k$. Parameter $B$ sets the vertical scale of the rotation curve
$v(x)$, 
so we can choose $B=1$ in what follows. The density profile $\tau(x)$
is a power law with index $k-1$, while $k$ can be thought as the power-law
index of the surface density profile $\sigma(x)\propto x^k$
\citep{chr07}.

Figs.~\ref{fig2}-\ref{fig4} show the shapes of the rotation curves obtained 
from eqs.~(\ref{rhoiso4}) and~(\ref{g}) for various choices of the
constants $A$ and $k$. The results are scale-invariant (a property of
power-law density profiles), so the radial scale is arbitrary (while, on the other
hand, $B=1$ scales $v$).
It is not surprising (see \S~\ref{physical} below)
that, in these Newtonian models, one sees most of
the shapes of the ``flat'' rotation curves observed in spiral galaxies.

The only shapes missing from the figures are those of the falling rotation
curves \citep{cas91}. Apparently, in some compact galaxies, 
the above equilibrium profiles did not endure (the sound-crossing time at 30~kpc for $C_o = 10$~km~s$^{-1}$ is 3~Gyr, so it seems that there has been enough time to achieve equilibrium); perhaps because
of interactions with nearby galaxies; or because the ``external'' gravity
of the massive bulges eliminates the intrinsic solution (see \S~\ref{physical}
and footnote~\ref{ft4} below). But even in these objects, the
falloff is slower than Keplerian which implies that gas pressure is still fighting to
establish its own preferred profiles. This must be the case
since the intrinsic solutions are favored by the equilibrium differential 
equation itself (see \S~\ref{math}) in the absence of other external forces. 
In the two galaxies observed by \cite{cas91},
certain segments of the rotation curves are flat and that indicates to us
that the radial self-gravitational equilibrium has fallen apart at some locations but not (yet) everywhere in these disks.

The above results can be summarized as follows: The derived density
profiles are simple power laws in radius and the rotation curves are flat or slightly
increasing at large radii (eqs.~[\ref{rhoiso4}] and~[\ref{g}]) irrespective 
of the value of the index $k$. Spiral galaxies have always been fitted
with exponential density profiles, thus we do not know which values of $k$
occur in nature. Galaxy profiles will have to be fitted again, but the payoff
this time will be substantial: when $k$ is determined from observations, the large-scale rotation curve (away from 
the center) will also be obtained independently from the velocity measurements.
Thus, the observational results will be tested for consistency within the same data set in each case. This also holds true for protoplanetary disks
in their early isothermal or adiabatic (see \S~\ref{poly} below) phases;
but first we need to find such purely self-gravitating disks in a 
pre-Class 0 YSO (Young Stellar Object)
stage, and the 21 cm HI line in emission or absorption may give us a chance \citep{kam08}.

\subsection{Physical Interpretation}\label{physical}

But how can such power-law density profiles produce and support
flat or slowly increasing rotation curves? The problem has 
always been that the centrifugal force remains too high in the outer regions
of the disk where Newtonian gravity weakens substantially. How do these
equilibrium models get around this discrepancy? The answer to these
questions was given in \cite{chr07} and we repeat it here:
The Lane-Emden eq.~(\ref{leiso1}) is a second-order differential 
equation. As such, it respects 
but does not rely solely on force balance.
(As will be readily seen in eq.~[\ref{balance1}] below, force balance 
is guaranteed by the way that the specific enthalpy of the gas is operating.)
This second-order equation describes locally the detailed radial 
($d/dx$) variation of
the {\it logarithmic gradients} ($d/d\ln x$) 
of the potentials involved in the struggle to reach equilibrium. 
So, initially, it is the log-gradient of the enthalpy (with help from rotation)
that sets out to oppose 
the log-gradient of the gravitational potential. This competition
can be seen by rewriting eq.~(\ref{leiso1}) in the form
\begin{equation}
\frac{1}{x}\frac{d}{dx}\left[\frac{d}{d\ln x}\left(h + \psi\right)\right] =
\frac{1}{x}\frac{d}{dx}\left[v^2\right] \ ,
\label{leiso2}
\end{equation}
where $h(x)\equiv\int{dp/\tau}$ and $\psi(x)$ are the dimensionless 
enthalpy per unit mass and the Newtonian self-gravitational potential, respectively.

And in that case, how and why does
the average intrinsic solution (eqs.~[\ref{rhoiso3}] and~[\ref{rhoiso4}]) come
into existence? The answer to these questions is even more illuminating:
The intrinsic solution was derived above by imposing two separate
conditions, that the centrifugal force should match 
the gravitational force (see eq.~[\ref{rhoiso1}] and the last two terms in eq.~[\ref{leiso2}]):
\begin{equation}
\frac{1}{x}\frac{d}{dx} \left(\frac{d\psi}{d\ln x}\right) = 
\frac{1}{x}\frac{d}{dx}\left(v^2\right) \ \Longrightarrow 
\ \frac{v^2}{x} = \frac{d\psi}{dx} \ ;
\label{balance1}
\end{equation}
while, at the same
time, the log-gradient of the enthalpy should retire
from the competition by assuming a constant profile
(see eq.~[\ref{rhoiso2}] and the leading term in eq.~[\ref{leiso2}]):
\begin{equation}
\frac{d}{dx} \left(\frac{dh}{d\ln x}\right) = 0 .
\label{balance2}
\end{equation}
This occurs only for a power-law density profile (eq.~[\ref{rhoiso3}]) 
and for a specific rotation profile (eqs.~[\ref{rhoiso4}] 
and~[\ref{g}]), and these profiles become internal properties characteristic
of the equilibrium disk. Stated more simply, up until now people believed
that rotation was not related to the structure of the equilibrium disk
and that they could adopt any arbitrary rotation profile for gaseous disks. 
We see now that this is not true, the radial density and rotation profiles are strongly coupled and uniquely determined through the intrinsic solution discussed above.

The only surprise in this narrative is the unique way that the differential
equation (or the disk) finds to promote and establish the above intrinsic solution: 
rather than trying to simultaneously balance the variations of the potentials involved at every single radius
(not possible because the corresponding timescales vary widely at 
different radii), the disk assumes gradually a
logarithmic specific-enthalpy profile determined from the solution of eq.~(\ref{balance2}):
\begin{equation}
\frac{dh}{dx} \propto \frac{1}{x} \ \Longrightarrow ~h(x)\propto \ln x~. 
\label{balance3}
\end{equation}
So it is the thermodynamic potential $h(x)$ of the gas that becomes logarithmic,
and not the gravitational potential that people have been trying to
make it so for nearly 50 years! Naturally, 
the disk establishes such a logarithmic profile because this law
guarantees precise force balance (eq.~[\ref{balance1}])
at all of its equatorial radii. The action of $h(x)$
unfolds in the physical disk from inside-out over timescales of the order of the local sound-crossing time $R/C_o$.
So the global equilibrium becomes complete after a time
$\sim R_{max}/C_o$, where $R_{max}$ is the outer radius of the disk. 

We understand physically the preceding results in the following manner:
Self-gravity is a long-range force (by Gauss's law, the gravitational 
potential at radius $x$ depends on the entire mass interior to $x$)
and it cannot adjust its potential 
in the disk to effect local changes to the density distribution.  In other words,
the density is a source term in the Poisson equation that determines the
gravitational potential, but not the other way around.
In stark contrast, enthalpy is a local potential
whose action depends only on the local behavior of the pressure and the 
density of the disk. When $h(x)$ assumes its logarithmic radial profile
(eq.~[\ref{balance3}]),
it dictates that the local density adjust according to the local
log-pressure gradient 
($\tau \propto dp/d\ln x $). By doing that, the enthalpy uses the density
in order to modify the sourcing of both the gravitational potential (via the Poisson equation $\nabla^2\psi = \tau$) and the centrifugal potential
(via eq.~[\ref{rhoiso1}]). The result of this tactic is a rotation law
that is entirely dependent on the distribution of enthalpy (see eq.~[\ref{rhoiso1}])
that does not feed back ($v$ does not enter in eq.~[\ref{rhoiso2}]).
The rotation so produced is capable of balancing gravity at all radii
all by itself (eq.~[\ref{balance1}]), and the enthalpy retires from
the struggle for equilibrium (eq.~[\ref{rhoiso2}]) having implicitly won 
the competition at every radius.

It is important to keep in mind that the enthalpy wins the struggle
because the two equations of the intrinsic equilibrium solution
(eqs.~[\ref{rhoiso1}],~[\ref{rhoiso2}] or eqs.~[\ref{balance1}],~[\ref{balance2}]) and the Poisson equation 
($\nabla^2\psi = \tau$) do not allow for feedback loops
and counter-sourcing by self-gravity or rotation. Thus,
the state of rotation, the density profile, 
and the local self-gravity have all
been manipulated unilaterally by the action of the enthalpy.

\subsection{Concluding Remarks}\label{iso_conclusion}

The above equations  give us a new probe into gaseous astrophysical
 disk systems (eqs.~[\ref{rhoiso1}] and~[\ref{rhoiso2}]). 
 For spiral galaxy disks, this probe
 ought to routinely confirm the above-described density-rotation coupling,
 as the observations already exist.
 At the same time, we are full of anticipation about what we can learn
 from the very early phases of purely self-gravitating protoplanetary disks 
 (before the central protostars form and dominate the dynamics)
 for which the rotation curves have not been measured with accuracy yet, but the radial density profiles in Class 0 YSOs have been obtained \citep{wil11}.
Our prediction, of course, is that, if found, such early disks 
(preferably earlier than Class 0) will be
observed to have ``flat'' rotation curves.\footnote{In the case of the
 Class 0 young system L1527 \citep{tob12,tob13}, the rotation curve is not flat,
 but we did not catch this 0.3~Myr old system early enough 
 (the sound-crossing time at 100~AU for a 10~K cold gas with 
 $C_o = 0.3$~km~s$^{-1}$ is 1500~yr).  On the other hand, \cite{yen15a,yen15b} report several other Class 0 YSOs whose rotation is not Keplerian outside of the inner few~AU.\label{ft3}}

\section{Polytropic Self-Gravitating Newtonian Gaseous Disks}\label{poly}

The cylindrical polytropic Lane-Emden equation \citep{lan69,emd07}
with rotation can be written in dimensionless form as
\begin{equation}
n c_o^2\cdot\frac{1}{x}\frac{d}{dx}x\frac{d}{dx}\tau^{1/n} + \tau =
\frac{1}{x}\frac{dv^2}{dx} \ ,
\label{lepoly1}
\end{equation}
where $n>0$ is the polytropic index and
the dimensionless constant sound speed $c_o$ was defined 
for $\rho = \rho_o$. (In general, the square of the sound speed 
$c^2(x)\equiv dp/d\tau$ varies as $\tau^{1/n}$ across the medium.)
This equation
describes the radial ($x$) equilibrium of a rotating, self-gravitating, 
gaseous disk or cylinder in which the gas obeys a polytropic equation
of state $p\propto\tau^{1+1/n}$. 
As in \S~\ref{iso}, eq.~(\ref{lepoly1}) is valid exactly for
infinite cylinders and to a high degree of approximation in the equatorial (symmetry) planes of disks (see the Appendix).
This latter point is supported by the calculations of \cite{schm87}, \cite{schm88}, and \cite{sch95} who studied also the stability
of thin-disk and cylindrical equilibria and found large regions of the parameter space with stable models for all values of $n > 1$; and a sizeable region
in which flattened disks with power-law density profiles were unstable to ring formation that causes their profiles to become oscillatory,
just as was predicted by the analysis of \S~\ref{math} above.

We repeat the procedure outlined in \S~\ref{iso} in order to obtain
the intrinsic solution of eq.~(\ref{lepoly1}): If we equate again the last two terms:
\begin{equation}
\tau(x) = \frac{1}{x}\frac{dv^2}{dx},
\label{rhopoly1}
\end{equation}
then this is an intrinsic solution provided that the rest of the equation 
(the radial variation of the logarithmic gradient of the enthalpy) 
vanishes:
\begin{equation}
\frac{d}{dx}x\frac{d}{dx}\tau^{1/n} = 0.
\label{rhopoly2}
\end{equation}
Eqs.~(\ref{rhopoly1}) and~(\ref{rhopoly2}) form a system
in which $v(x)$ is totally dependent on $\tau(x)$.  
First we solve eq.~(\ref{rhopoly2}) to
obtain the radial density profile:
\begin{equation}
\tau(x) = \left[\ln (A x^k)\right]^n, \ \ (A, k = {\rm const.}),
\label{rhopoly3}
\end{equation}
and then we solve eq.~(\ref{rhopoly1}) to determine the rotation curve
of the intrinsic solution:
\begin{equation}
v(x) = \sqrt{\frac{A^{-2/k}}{2} \left(\frac{-k}{2}\right)^n \cdot 
\Gamma\left(n+1, \ln\frac{A^{-2/k}}{x^2}\right) + B}, \ \ (B = {\rm const.}),
\label{rhopoly4}
\end{equation}
where $A>0$, $n>0$, $k<0$, $A x^k \geq 1$ (i.e., $x\leq A^{-1/k}$),  
and the upper incomplete gamma function is defined as
\begin{equation}
\Gamma (\alpha, z) \equiv \int_z^\infty{e^{-t} t^{\alpha - 1} dt } \ ,
\ \ (z\geq 0) \ .
\label{gamma}
\end{equation}
The solution contains 4 free parameters, the integration constants
$A$, $B$, $k$, and the polytropic index $n$. Parameter $B$ sets the vertical scale of the rotation curve
$v(x)$, so we can choose $B=1$ in what follows. 

Figs.~\ref{fig5}-\ref{fig8} show the shapes of the rotation curves obtained 
from eq.~(\ref{rhopoly4}) for two polytropes with $n=1.5$ and $n=3$ 
and for various choices of the constants $A$ and $k$. 
As in the isothermal case of \S~\ref{iso}, the rotation profiles are
slowly increasing or flat with radius $x$. In this case however, we need to obey
the condition $x\leq A^{-1/k}$ ($z\geq 0$ in eq.~[\ref{gamma}]) in the calculation of the gamma function
and so the rotation curves terminate when $x$ reaches its maximum value.
Two basic trends are noted in the figure captions as well: (a)~for fixed $n$, the curves rise more steeply for steeper values of the index $k<0$;
and (b)~for fixed $k$, the curves become flatter for higher values 
of the polytropic index $n>0$.

\subsection{Physical Interpretation}\label{physical2}

The polytropic Lane-Emden equation with rotation and its intrinsic solution
assume the exact same forms as in the isothermal case (\S~\ref{physical})
when the polytropic equation of state $p\propto\tau^{1+1/n}$ is used to introduce the specific enthalpy $h(x)\equiv\int{dp/\tau}$. Therefore, the fundamental equations discussed in \S~\ref{physical} also apply to polytropic models with finite radial extent and the enthalpy plays the exact same role in manipulating the source terms of the gravitational potential and the rotational potential.

The fact that the thermodynamic potential $h(x)$ operates locally 
in the exact same manner (i.e., $h\propto \ln x$) in polytropic and isothermal equilibria 
helps us correct another common misconception:
Since the time that the results of \cite{hay82} came to light
(they studied isothermal self-gravitating gaseous disks that oddly exhibited power-law density profiles and ``flat'' rotation curves in equilibrium),
it has been often stated that isothermal disk models tend to exhibit
flat rotation curves because they happen to have a ``special'' mass
distribution (i.e., their specific angular momentum is proportional to the
mass interior to radius $x$, or their mass grows linearly with $x$). 
This is not true. The above results show without a doubt that
there are no special equilibrium models; and that the isothermal and
the polytropic equilibria
are both subject to the same fundamental physics at the local level, where
the thermodynamic potential $h(x)$ operates and dominates when the only
gravity it faces is the self-gravity of the disk. As for the validity 
of using the cylindrical coordinate system with its ``special'' $\nabla^2$ operator, this issue is addressed in detail in the Appendix.

Furthermore, as we have seen above, a flat rotation curve does not imply
and does not need a linearly increasing mass distribution (see also the
Appendix).
Even if the underlying density profile is steeply decreasing in radius (as the light is in
spiral galaxy disks), a ``flat'' rotation curve is a requirement in the equilibrium solutions derived in this section and in \S~\ref{iso}.
Therefore, one can assume a constant mass-to-light ratio and build a Newtonian self-gravitating galaxy disk model with a steeply declining light distribution that will still be required to have a rising rotation curve.

\subsection{Concluding Remarks}\label{poly_conclusion}

The polytropic intrinsic solution (eqs.~[\ref{rhopoly1}] and~[\ref{rhopoly2}]) gives us a new probe into nonisothermal astrophysical disk systems. 
In particular, protoplanetary disks undergo various early phases of adiabatic
evolution \citep{toh02}. We predict that, if found, such disks
will be observed to have the same fundamental characteristics
(eqs.~[\ref{rhopoly3}] and~[\ref{rhopoly4}]) irrespective of the
polytropic index $n$ appropriate for each adiabatic phase.
In fact, observations of the density profiles, fitted with
eq.~(\ref{rhopoly3}), may be able to determine,
not only the profile constants $k$ and $A$, but also the value of $n$,
thereby deriving the equation of state of the gas independently from theoretical models.
The only problem is that we need to find such systems very early in their
development (see footnote~\ref{ft3} above), and this is a very difficult task
\citep{kam08,wil11,tsi13}.

\section{Discussion}\label{discussion}

In this paper, we have investigated the Lane-Emden equations with rotation
that describe the equilibrium structures of rotating, self-gravitating, gaseous disks and cylinders (\S~\ref{iso} and~\ref{poly}; see also the Appendix). 
We have obtained new analytic singular solutions that we call intrinsic solutions because they are dictated and favored by the differential
equations themselves with no regard to physical boundary conditions
that are externally imposed and that shape up the regular solutions of
the equations (the Cauchy problem). In \S\S~\ref{math}-\ref{poly}, we have effectively shown that second-order differential equations (both linear and nonlinear) are at
odds with the Cauchy problem because the equations show a strong preference for
their own intrinsic solutions that, for inhomogeneous equations
of the Lane-Emden type, cannot be obtained by solving the 
boundary-value
problem (hence they are singular). 
The regular Cauchy-type solutions are then attracted to and forced to oscillate about the
intrinsic solutions, which means that they do their best to match those
dominant solutions. This
results in ``regular'' oscillatory density and rotation profiles whose averages 
are precisely the underlying intrinsic solutions \citep{chr07,chr15}. 
In this sense, the differential
equations succeed in imposing their preferences to the Cauchy problem.
This, by itself, is an important conclusion that has ramifications beyond astrophysics for the theory of second-order differential equations
of mathematical physics.

The intrinsic solutions are very much related to the so-called trivial solutions
of differential equations \citep{chr07}. We now understand that there are no trivial
solutions, in fact such solutions of second-order equations are quite dominant
(see \S~\ref{math}): In many cases of interest, knowing the trivial solution of an equation implies that
we know the average behavior of all the regular solutions that depend on various types of boundary conditions but, nevertheless, end up oscillating about the intrinsic solution, provided that the differential equation
is a harmonic oscillator (as the ordinary Bessel equations in \S~\ref{math};
the nonisothermal Lane-Emden equations without rotation in \S~\ref{math};
and the Lane-Emden equations with rotation in \S\S~\ref{iso}-\ref{poly}). 

The mean density and rotation profiles that we have derived analytically
in \S\S~\ref{iso} and~\ref{poly} are dominated by natural logarithms
and power laws.
This is the implicit reason that \cite{mar15} has recently suceeded in matching
the shapes of the rotation curves of a sample of 37 spiral galaxies by 
using the log-normal probability distribution (and no dark matter or modified gravity)
to describe the density profiles in the equatorial planes of the disks. This surface density distribution (eq.~[2] in Marr 2015) is equivalent to a variable power law of the form $x^{k(x)}$ in normalized radius $x$ in which the index $k(x)$ varies across the disk as
\begin{equation}
k(x) = -1 - \frac{1}{2 s^2}\ln x  \ ,
\label{marr_k}
\end{equation}
where $s^2$ is the variance of the distribution, a free parameter to be fitted for each spiral galaxy model.
In principle, a slowly varying $k(x)$ is permitted by our analytic solution because the specific enthalpy $h(x)$ is a strictly local potential function (eq.~[\ref{balance2}] in \S~\ref{physical}); and if the power-law index $k$ has to vary radially in order for the
equilibrium disk to obey some other fundamental law (e.g., as Marr states, the total entropy of the overall configuration should be maximized), then such adjustment 
may occur on timescales determined by the local sound speed.

Returning to the astrophysical context, the intrinsic solutions of
the various types of the Lane-Emden equation with rotation have, for the first time,
succeeded in explaining the flat rotation curves of spiral galaxy disks
without the need of invoking dark matter or modified gravity.
Flat rotation curves are a rigorous requirement of Newtonian gravity
in gaseous self-gravitating astrophysical disks.
The only fair way to describe this result (see \S\S~\ref{iso},~\ref{poly}) is that Sir Isaac \cite{new87} is vindicated 
once again, and our searches for dark matter in the universe 
and our attempts to modify Newtonian gravity (references are listed in \S~\ref{intro}) have sadly amounted to just a ``wild-goose chase.'' 
In fact, since the flat rotation curves of spiral galaxies can now be
explained at such a fundamental level, the massive observational results
collected over the years 
must be considered as yet another test that Newtonian gravity has successfully passed on scales of $\sim 10-100$~kpc and at nonrelativistic velocities. This is an impressive
achievement when compared to previous tests conducted on and limited to scales
no larger than that of our solar system.

Our results render the Dark Matter Hypothesis unnecessary on galaxy scales.
This removes the largest pillar of this hypothesis (flat rotation
curves have remained to this day the ``strongest piece of evidence'' in favor
of dark matter, but not any longer). 
But this ``aetherial'' hypothesis is not about to roll over and die
without a fight. 
The next areas of confrontation will be larger than galaxy scales
and cosmology. We are very much encouraged from a recent report of the 
absence of dark matter on larger than galaxy-disk scales: \cite{mag13} 
analyzed 25 gravitational lenses and found that their mass determinations
indicate the absence of extended dark matter haloes all the way out to
distances comparable to the Einstein ring (the separation between lensed
quasar images). 

On the other hand, we do not anticipate any serious problems
materializing in cosmology because we do not believe that there currently is any credible observational evidence in favor of dark matter or modified gravity on those largest scales. In fact, some results that argue against the necessity for dark matter on various scales have timidly begun to appear 
\citep{lop15,lan15,mon12,mon15}.
For these reasons, here is how we approach the issue of dark matter now:
Christiaan Huygens presented his theory of elastic longitudinal light waves
propagating in ``aether'' to the Paris Academy of Sciences
in 1678 and published his views a few years
later \citep{huy90}. That year, the physics world entered a Dark Age
that lasted nearly 200 years, until the genious of \cite{mm87} finally showed 
that the universe is not filled with aether.  It seems that we also live 
in another Dark Age, the ``Dark Matter Dark Age,'' that commenced in the 1970s when K. C.~\cite{fre70} and others reported that the rotation curves of spiral galaxies were not falling with radius. By all accounts, the current Dark Age
has lasted for nearly 50 years; and the sooner we get out of it, the better
for our understanding of the large scales of the universe around us.

The analytic solutions that we have derived in this work should also
find applications in the field of protoplanetary disk research 
(\S\S~\ref{iso_conclusion} and~\ref{poly_conclusion}), especially
if very young, purely self-gravitating disks could be found in the future
\citep{kam08}.
At present, only observations of Class 0 YSOs are widely available \citep{wil11}, 
but these systems are not young enough and do not have 
flat rotation curves; their protostars have formed
and they are changing the dynamics and kinematics 
of the disks.\footnote{When a protostar grows at the center of a protoplanetary disk, the enthalpy is defeated by gravity because
the enthalpy cannot source and manipulate this component of the gravitational field
via the Poisson equation (see \S~\ref{physical});
and the intrinsic solutions described in this work are no longer valid
for such gaseous disks subject to ``external'' gravitational fields.\label{ft4}}
We are encouraged however by the report of \cite{tsi13} who discovered
an increasing rotation curve in the inner $2000-8000$~AU of the ``first hydrostatic core'' candidate Cha-MMS1. 

\acknowledgments

We thank Joel Tohline for feedback and guidance over many years;
and John Marr, Hsi-Wei Yen, and Earl Schulz for many fruitful discussions, especially those concerning the observations of astrophysical self-gravitating disks. DMC was supported in part by NASA grant NNX14-AF77G.



\appendix
\section{Vertically Thin Disks Versus Infinite Cylinders}

When applied to the equatorial planes ($Z=0$) of thin disks, eqs.~(\ref{leiso1}) and~(\ref{lepoly1}) do not include the specific enthalpy term
\begin{equation}
\frac{d^2h}{dz^2}\left(x, z=0\right) \ , 
\label{z1}
\end{equation}
where $z\equiv Z/R_o$. This term is small for cold gases since it scales as the sound speed squared $c_o^2$. Nevertheless, strong objections have been raised about the validity of this approximation. Here we address such objections as follows.

The analysis presented in \S\S~\ref{iso} and~\ref{poly} shows that the
specific enthalpy in cylinders assumes a logarithmic radial profile 
(eq.~[\ref{balance3}]). Now, pressure is an isotropic force and its nature is to push spherically out in self-gravitating gases.
Therefore, eq.~(\ref{balance3}) suggests that,
in axisymmetry, the specific enthalpy of the gas ought to assume a spherical form
\begin{equation}
h(r) \ \propto \ \ln r \ = \ \frac{1}{2}\ln(x^2 + z^2)\ , 
\label{z2}
\end{equation}
where $r$ is the spherical radius normalized by $R_o$.
This behavior is artificially suppressed in cylinders by dropping the
dependence on $z$ from the equations. But, in principle, it cannot be suppressed
for thin disks, so this term should at least be quantified in the equatorial planes of disks:

(a)~Applying the radial ($x$) component of the cylindrical Laplacian in eq.~(\ref{z2})
and then setting $z=0$, we confirm eq.~(\ref{balance3}); that is,
this term of the Lane-Emden equations vanishes
on the equatorial plane,
as was found in the intrinsic solutions of \S\S~\ref{iso} and~\ref{poly}.

(b)~Combining eqs.~(\ref{z1}) and~(\ref{z2}), we find that
\begin{equation}
\frac{d^2h}{dz^2}\left(x, z=0\right) \ \propto \ \frac{1}{x^2} . 
\label{z3}
\end{equation}
This is the term that was ignored in the force balance described by
eq.~(\ref{balance1}). But it makes only a minor contribution
to the force balance over all radii where $c_o << v(x)$. Specifically,
it modifies eqs.~(\ref{rhoiso1}) and~(\ref{rhopoly1}) to
\begin{equation}
\tau(x) = \frac{1}{x}\frac{dv^2}{dx} - \frac{\ell}{x^2} \ ,
\label{z4}
\end{equation}
where $\ell < 0$ is a proportionality constant of order $c_o^2$.
For our models, $\ell = (k-1)c_o^2$ for isothermal disks 
and $\ell = n k c_o^2$ for polytropic disks.
Since $\ell < 0$ (because $k<0$), the last term in eq.~(\ref{z4}) opposes self-gravity.
By integration of this equation, we find that its contribution to
the rotation profile $v^2(x)$ is logarithmic:
\begin{equation}
v^2(x) = \int{\tau(x) x dx} + \ell\ln x  + B\ ,
\label{z5}
\end{equation}
where $B$ is the integration constant.
We see now that $B$ and the dimensionless cylindrical ``mass'' function
$m(x) = \int{\tau x dx}$ make the largest contributions to the rotation curve. 
This dependence ($v^2\propto B + m$) differs conclusively from the conventional
thinking that is responsible for the acceptance of dark matter; that 
$v^2/r = G m/r^2$, thus a constant $v$ requires $m\propto r$.

\begin{figure}
\epsscale{.80}
\plotone{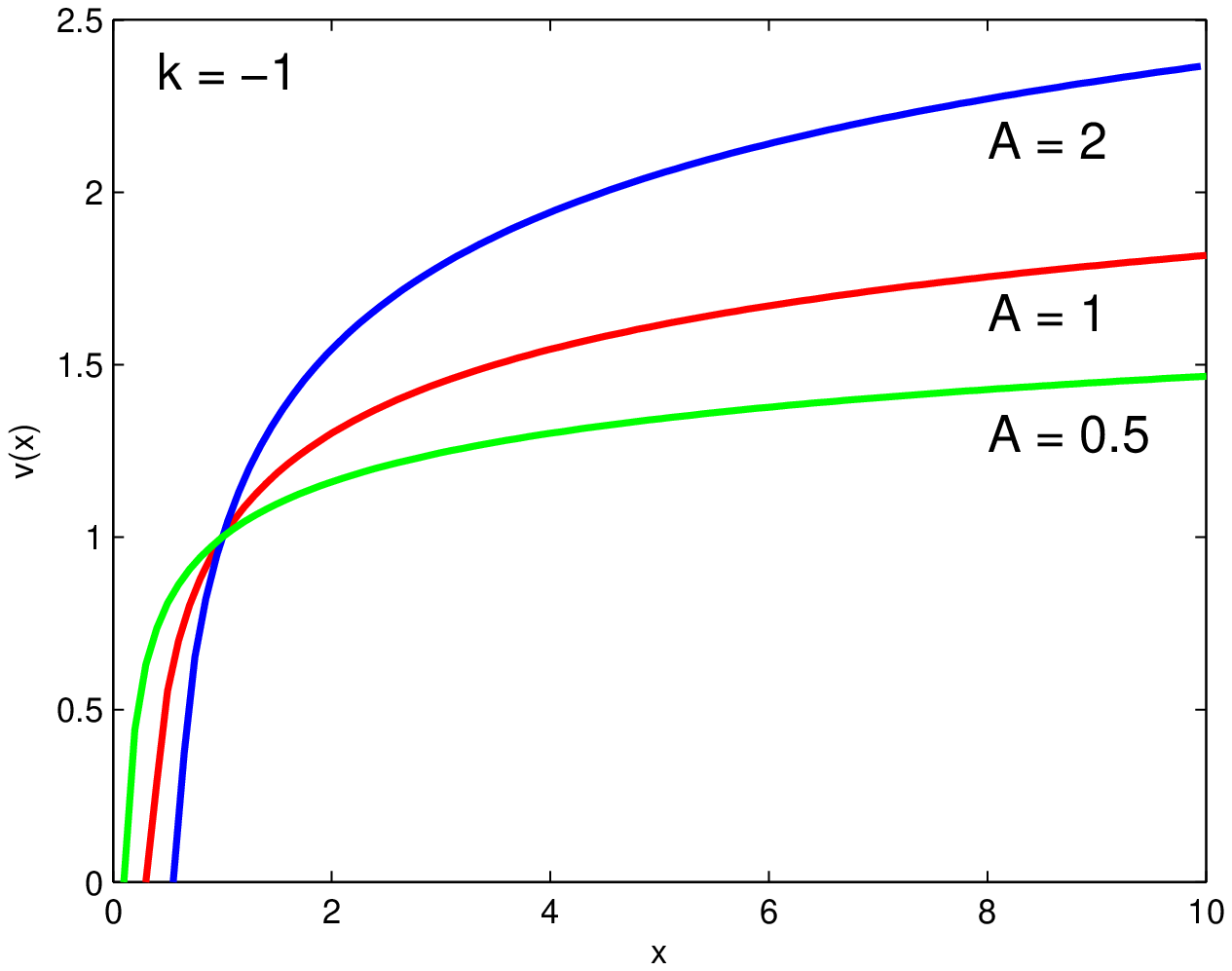}
\caption{Rotation curves of the intrinsic solution of the isothermal
Lane-Emden equation for $k=-1$, $B=1$, and various values of $A$.
\label{fig2}}
\end{figure}

We should elaborate on the $B + m$ term in the above analysis. 
Hidden in this term are
the two different Newtonian models that we have discovered and that require rising rotation curves in disks and cylinders:

(1)~In all isothermal models with $k\geq -1$ (Fig.~\ref{fig2})
and in all polytropic models (Figs.~\ref{fig5}-\ref{fig8}), the indefinite integral in eq.~(\ref{z5}) is positive definite.
Then $v^2\propto B + m(x)$, and the rotation curves rise at all radii as more mass is added by the integration.

\begin{figure}
\epsscale{.80}
\plotone{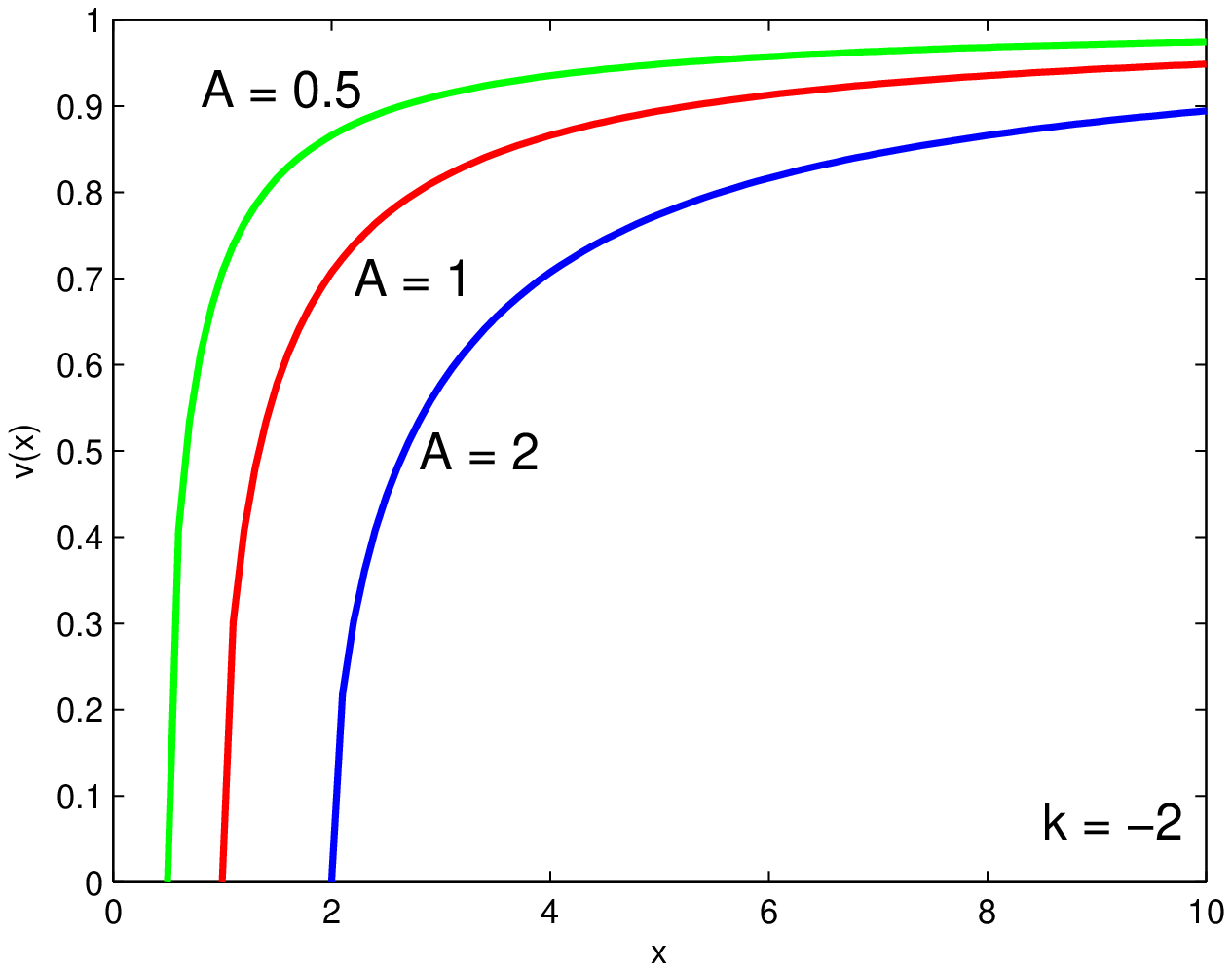}
\caption{Rotation curves of the intrinsic solution of the isothermal
Lane-Emden equation for $k=-2$, $B=1$, and various values of $A$.
\label{fig3}}
\end{figure}

\begin{figure}
\epsscale{.80}
\plotone{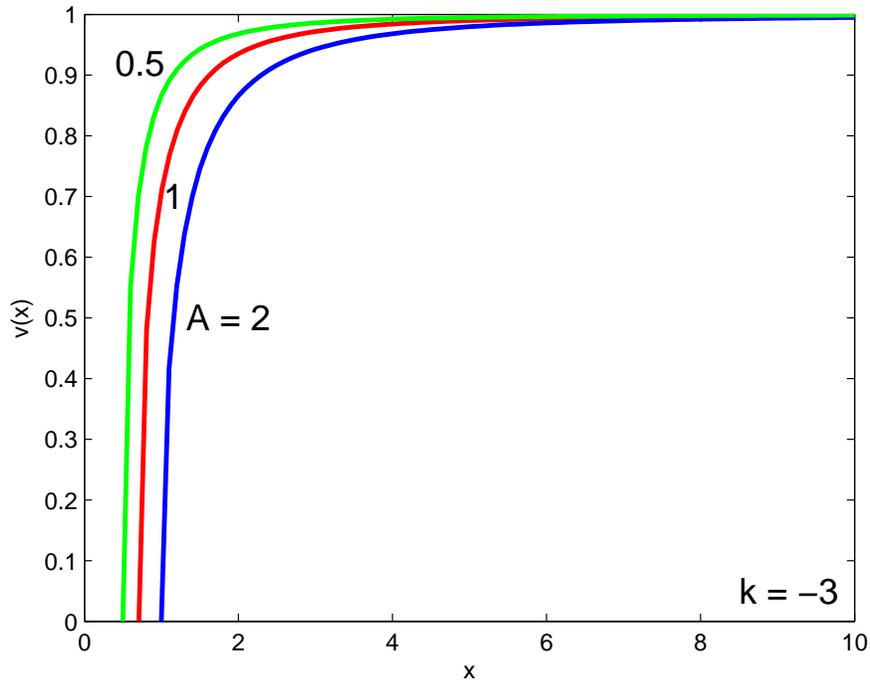}
\caption{Rotation curves of the intrinsic solution of the isothermal
Lane-Emden equation for $k=-3$, $B=1$, and various values of $A$. 
Figs.~\ref{fig2}-\ref{fig4} show that
the curves become flatter for steeper values 
of the power-law index $k$.
\label{fig4}}
\end{figure}

\begin{figure}
\epsscale{.80}
\plotone{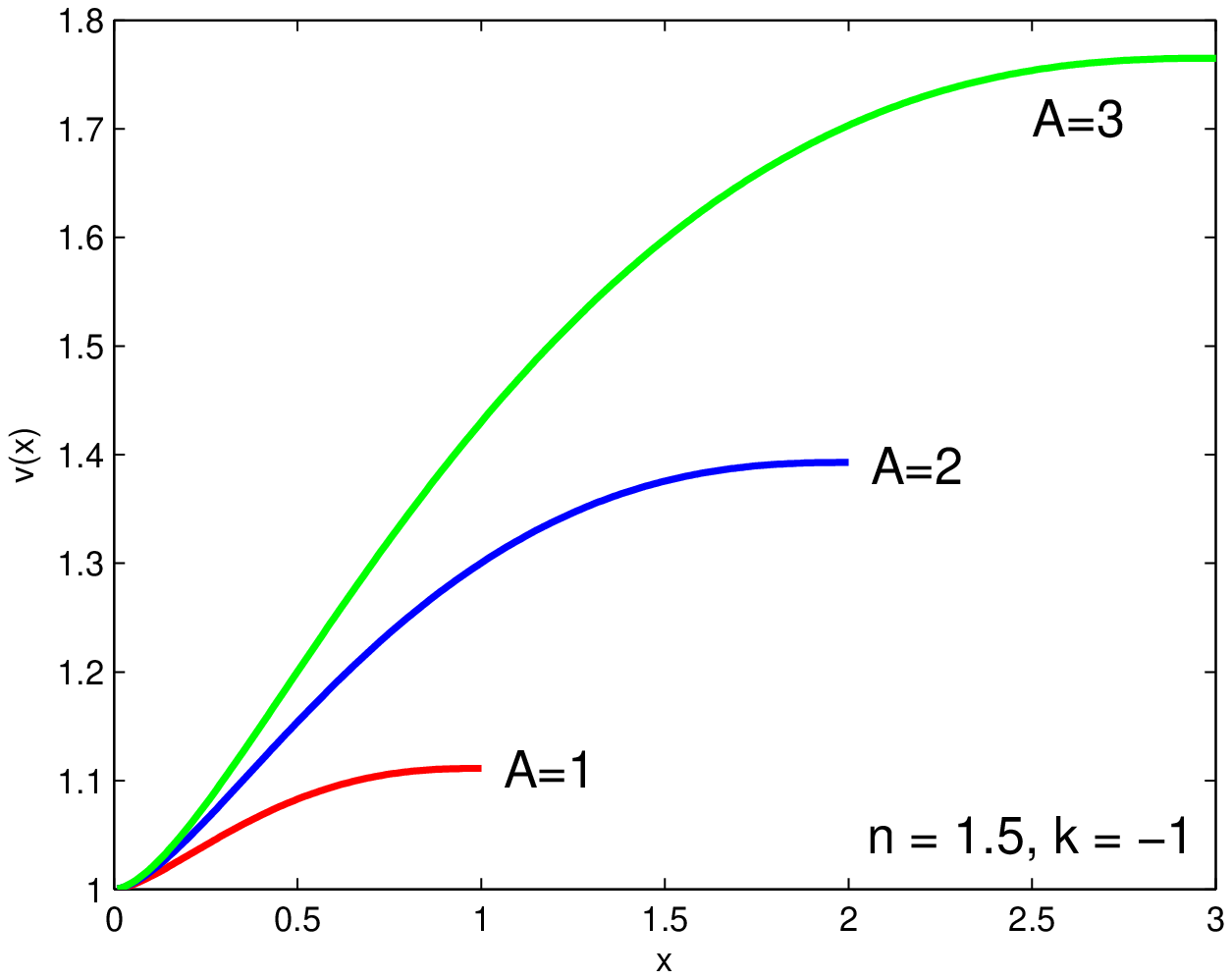}
\caption{Rotation curves of the intrinsic solution of the $n=1.5$ polytropic
Lane-Emden equation for $k=-1$, $B=1$, and various values of $A$.
\label{fig5}}
\end{figure}

\begin{figure}
\epsscale{.80}
\plotone{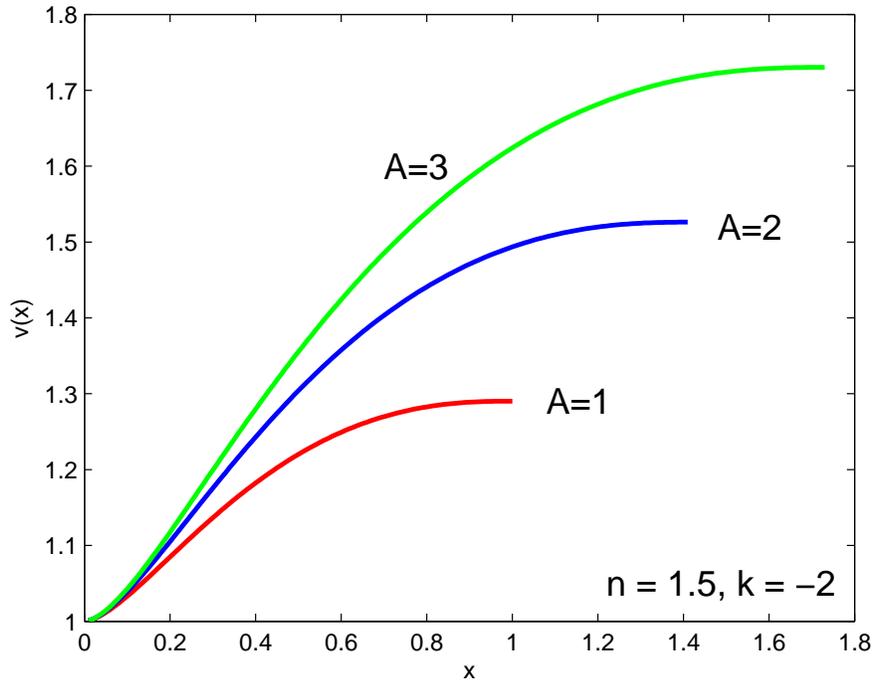}
\caption{Rotation curves of the intrinsic solution of the $n=1.5$ polytropic
Lane-Emden equation for $k=-2$, $B=1$, and various values of $A$.
Figs.~\ref{fig5} and~\ref{fig6} show that,
for fixed $n$, the curves rise more steeply for steeper values 
of the index $k$.
\label{fig6}}
\end{figure}

\begin{figure}
\epsscale{.80}
\plotone{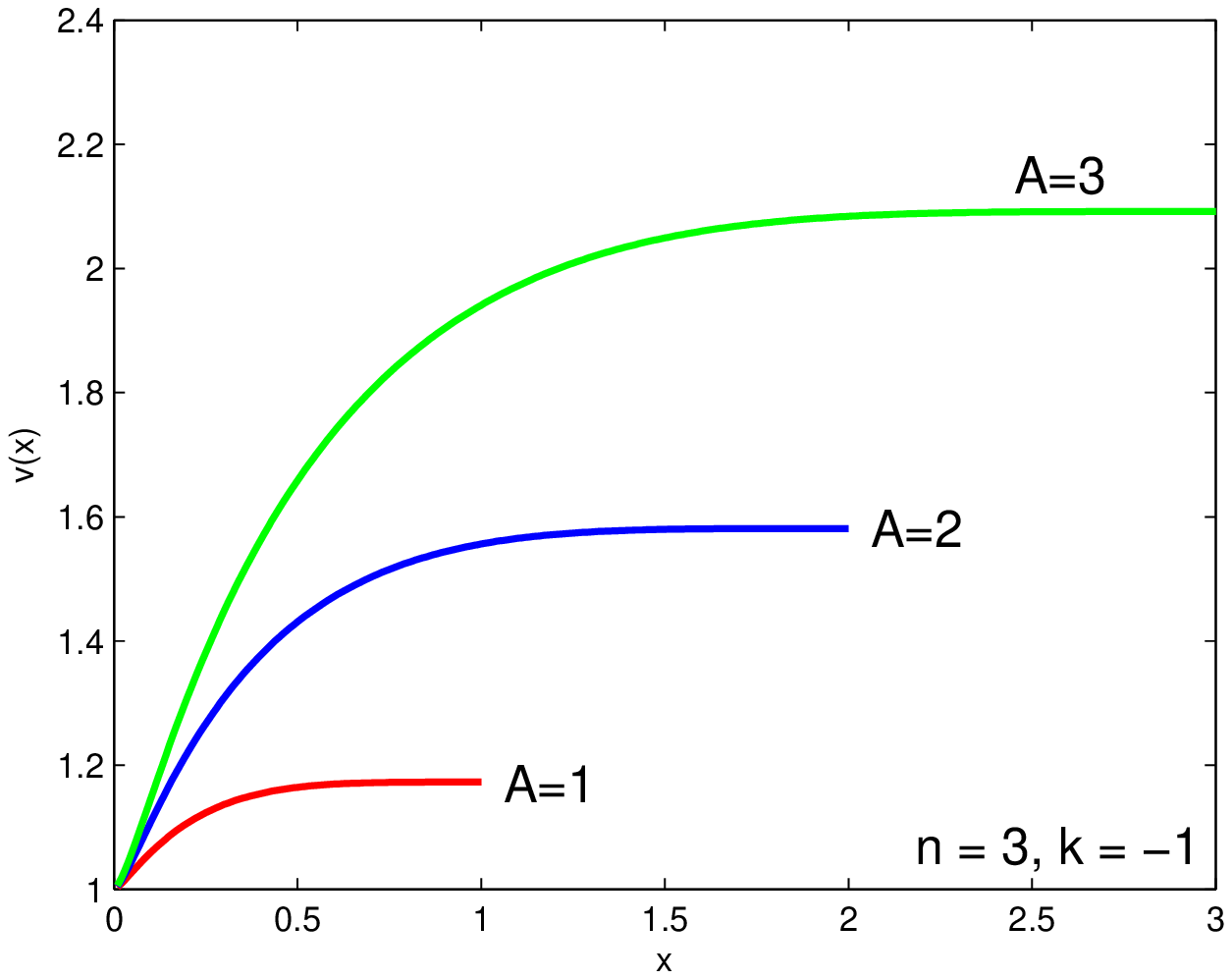}
\caption{Rotation curves of the intrinsic solution of the $n=3$ polytropic
Lane-Emden equation for $k=-1$, $B=1$, and various values of $A$.
\label{fig7}}
\end{figure}

\begin{figure}
\epsscale{.80}
\plotone{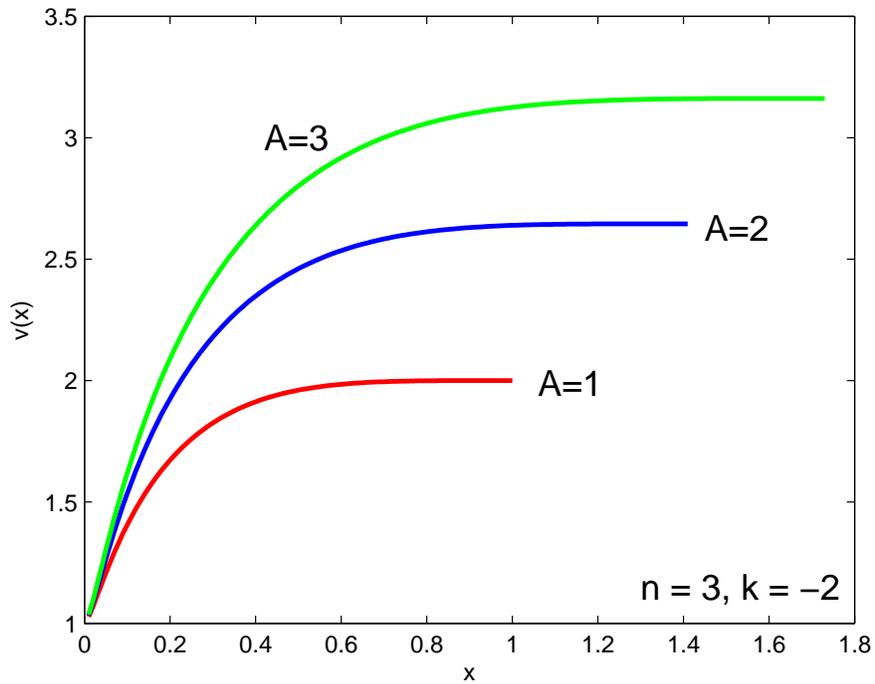}
\caption{Rotation curves of the intrinsic solution of the $n=3$ polytropic
Lane-Emden equation for $k=-2$, $B=1$, and various values of $A$.
Figs.~\ref{fig6} and~\ref{fig8} show that,
for fixed $k$, the curves become flatter for higher values 
of the polytropic index $n$.
\label{fig8}}
\end{figure}

(2)~On the other hand, in isothermal models with $k < -1$ 
(Figs.~\ref{fig3} and~\ref{fig4}), the indefinite intergal in eq.~(\ref{z5}) is negative definite (although the definite integral for the mass is still positive). 
Then, in effect, $v^2\propto B - |m(x)|$, and the rotation curves approach asymptotically the constant value $v=\sqrt{B}$ as $m(x)$ decreases 
with $x$. Because $\tau(x)$ is a very steeply decreasing function of $x$ in such cases,
then $m(x)\to 0$ from below quite fast, and that makes the rotation curves appear quite flat  over most radii in Figs.~\ref{fig3} and~\ref{fig4}.

Returning to the pressure term ($\ell\ln x$) in eq.~(\ref{z5}),
if it is included in $v^2(x)$ in future models, this term 
has the potential to drive the rotation curves down at large radii
because it is negative for $x>1$.
This behavior can occur only in disks since this term
has been suppressed in infinite cylinders. So, unlike disks, cylindrical
filaments can 
{\it never} exhibit falling rotation curves in equilibrium
(unless of course their age is smaller than the sound-crossing time).
Our analysis in \S\S~\ref{iso} and~\ref{poly} was carried out
without this term in order to bring out the physics of such Newtonian
systems. Nevertheless, the pressure term in eq.~(\ref{z5}) or some similar approximation may be of interest
to researchers planning to remodel the rotation curves of spiral galaxies.
But it does not appear to be necessary to modeling the rotation of
pre-Class 0 protoplanetary disks because such starless disks are
subject to extended infall of matter that ought to create cylindrical
quasi-equilibria to a good approximation.





\clearpage

\clearpage

\clearpage

\clearpage

\clearpage

\clearpage

\clearpage

\clearpage

\clearpage

\clearpage

\end{document}